\documentclass[twocolumn,trackchanges, resetfootnote]{aastex7}

\usepackage{booktabs}
\usepackage{filecontents,catchfile}
\usepackage{longtable}
\usepackage{cleveref}
\usepackage{CJKutf8}
\usepackage{footnote}

\newcommand{\RN}[1]{%
  \textup{\uppercase\expandafter{\romannumeral#1}}%
}

\newcommand{\um}{\,\micron}
\newcommand{\NeII}{[Ne\,\textsc{ii}]}
\newcommand{\NeIII}{[Ne\,\textsc{iii}]}

\newcommand{\OIV}{[O\,\textsc{iv}]}
\newcommand{\FeII}{[Fe\,\textsc{ii}]}
\newcommand{\ArII}{[Ar\,\textsc{ii}]}

\newcommand{\molH}{H\ensuremath{_\mathrm{2}}}
\newcommand{\HII}{H\,\textsc{ii}}

\newcommand{\JWST}{\textit{JWST}}

\newcommand{\target}{II~Zw~40}
\newcommand{\Pfdelta}{\mathrm{Pf}\,\delta}
\newcommand{\Gnought}{G{$_\mathrm{0}$}}

\renewcommand{\thefootnote}{\fnsymbol{footnote}}
\shorttitle{Resolving PAHs in blue compact dwarf II Zw 40}
\shortauthors{Lai et al.}

\begin{document}
\begin{CJK*}{UTF8}{bsmi}
\title{Resolving Emission from Small Dust Grains in the Blue Compact Dwarf II~Zw~40 with JWST}

\correspondingauthor{Thomas S.-Y. Lai}

\author[0000-0001-8490-6632]{Thomas S.-Y. Lai (賴劭愉)}
\affil{IPAC, California Institute of Technology, 1200 E. California Blvd., Pasadena, CA 91125, USA}
\email[show]{ThomasLai.astro@gmail.com}

\author[0000-0003-0014-0508]{Sara Duval}
\affiliation{Ritter Astrophysical Research Center, University of Toledo, Toledo, OH 43606, USA}
\email{}

\author[0000-0003-1545-5078]{J.D.T. Smith}
\affiliation{Ritter Astrophysical Research Center, University of Toledo, Toledo, OH 43606, USA}
\email{}

\author[0000-0003-3498-2973]{Lee Armus}
\affil{IPAC, California Institute of Technology, 1200 E. California Blvd., Pasadena, CA 91125, USA}
\email{}

\author[0000-0003-0760-4483]{Adolf N. Witt}
\affiliation{Ritter Astrophysical Research Center, University of Toledo, Toledo, OH 43606, USA}
\email{adolf.witt@utoledo.edu}

\author[0000-0002-4378-8534]{Karin Sandstrom}
\affiliation{Department of Astronomy \& Astrophysics, University of California, San Diego, 9500 Gilman Drive, La Jolla, CA 92093}
\email{kmsandstrom@ucsd.edu}

\author[0000-0003-1356-1096]{Elizabeth Tarantino}
\affiliation{Space Telescope Science Institute, 3700 San Martin Drive, Baltimore, MD 21218, USA}
\email{}

\author[0000-0002-9850-6290]{Shunsuke Baba}
\affiliation{Institute of Space and Astronautical Science, Japan Aerospace Exploration Agency, 3-1-1 Yoshinodai, Chuo-ku, Sagamihara, Kanagawa 252-5210, Japan}
\email{s-baba@ir.isas.jaxa.jp}

\author[0000-0002-5480-5686]{Alberto Bolatto}
\affiliation{Department of Astronomy, University of Maryland, College Park, MD 20742, USA}
\email{}

\author[0009-0001-6065-0414]{Grant P. Donnelly}
\affil{Ritter Astrophysical Research Center, University of Toledo, Toledo, OH 43606, USA}
\email{}

\author[0000-0001-7449-4638]{Brandon S. Hensley}
\affiliation{Jet Propulsion Laboratory, California Institute of Technology, 4800 Oak Grove Drive, Pasadena, CA 91109, USA}
\email{}

\author[0000-0001-6186-8792]{Masatoshi Imanishi}
\affiliation{National Astronomical Observatory of Japan,
National Institutes of Natural Sciences (NINS), 2-21-1 Osawa,
Mitaka, Tokyo 181-8588, Japan}
\affiliation{Department of Astronomy, School of Science,
Graduate University for Advanced Studies (SOKENDAI), Mitaka,
Tokyo 181-8588, Japan}
\email{masa.imanishi@nao.ac.jp}

\author[0000-0003-4023-8657]{Laura Lenkic}
\affiliation{IPAC, California Institute of Technology, 1200 E. California Blvd., Pasadena, CA 91125, USA}
\email{}

\author[0000-0002-1000-6081]{Sean Linden}
\affiliation{Steward Observatory, University of Arizona, 933 N Cherry Avenue, Tucson, AZ 85721, USA}
\email{}

\author[0000-0002-6660-9375]{Takao Nakagawa}
\affiliation{Institute of Space and Astronautical Science, Japan Aerospace Exploration Agency, Chuo-ku, Sagamihara, Kanagawa 252-5210, Japan}
\affiliation{Advanced Research Laboratories, Tokyo City University, Setagaya-ku, Tokyo 158-8557, Japan}
\email{nakagawa@ir.isas.jaxa.jp}

\author[0000-0002-8712-369X]{Henrik W.W. Spoon}
\affiliation{CCAPS, Cornell University, Ithaca, NY 14853, USA}
\email{}

\author[0000-0001-5042-3041]{Aditya Togi}
\affiliation{Texas State University, 601 University Dr, San Marcos, TX 78666, USA}
\email{}

\author[0000-0003-2093-4452]{Cory M. Whitcomb}
\affil{Ritter Astrophysical Research Center, University of Toledo, Toledo, OH 43606, USA}
\email{}

\begin{abstract}
 We present James Webb Space Telescope (JWST) Near Infrared Spectrograph (NIRSpec) and Mid-infrared Instrument (MIRI) integral-field spectroscopy of the nearby blue compact dwarf \target, which has a low metallicity of 25\% of solar. Leveraging the high spatial/spectral resolution and wavelength coverage of JWST/NIRSpec, we present robust detections of the 3.3\um\ polycyclic aromatic hydrocarbon (PAH) emission on 20~pc scales. The strength of the $\Pfdelta$ emission relative to the 3.3 PAH feature is significantly stronger than typical higher metallicity star-forming galaxies. We find that 3.3\um\ PAH emission is concentrated near the northern super star cluster and is co-spatial with CO gas. A strong correlation exists between the 3.3/11.3 PAH ratio and radiation hardness probed by \NeIII/\NeII, providing evidence of photodestruction of PAH molecules in intense radiation environments. Our analysis shows that while the overall PAH fraction is lower in \target\ than in higher metallicity galaxies, the contribution of the 3.3\um\ PAH feature to the total PAH emission is higher. We propose that the PAH size distribution is fundamentally shaped by two competing mechanisms in low-metallicity environments: photo-destruction and inhibited growth. Additionally, the high radiation field intensity in \target\ suggests that multi-photon heating of PAHs may be an important effect. As one of the first spatially resolved studies of aromatic emission in a low-metallicity environment, our spectroscopic results offer practical guidance for future observations of the 3.3\um\ PAH feature in low-metallicity galaxies using JWST.
\end{abstract}

\keywords{Blue compact dwarf galaxies (165) --- Interstellar medium (847) --- Polycyclic aromatic hydrocarbons (1280) --- Infrared spectroscopy (2285)}


\section{Introduction} \label{sec:intro}
Blue compact dwarf (BCD) galaxies constitute the least chemically evolved star-forming systems in the local universe \citep{Thuan2008}, making them excellent laboratories for examining primordial star formation and the build-up of metal content in the early universe. BCDs are characterized by their low luminosity (M$_\mathrm{V}\gtrsim-18$), compact size ($\leq$1 kpc in diameter), blue optical color, and low metal abundance, with oxygen abundances 12+log(O/H) ranging from 7.1 to 8.3 ($\sim$3---30\% Z$_{\odot}$) \citep{Thuan1981}.

The ISM conditions in metal-poor galaxies and typical starburst galaxies exhibit significant differences \citep[e.g.][]{Henkel2022}. Between galaxies with extremely low metallicity (2\% Z$_{\odot}$) and those approaching solar metallicity, the gas-to-dust mass ratio differs by three orders of magnitude \citep{Remy-Ruyer2014}, and the dust-to-metal ratio varies by about an order of magnitude \citep{DeVis2019}. In low-metallicity galaxies, \citet{Remy-Ruyer2015} identified far-infrared dust SEDs with broader profiles and peaks at shorter wavelengths compared to their metal-rich counterparts. Another notable difference is the reduction in polycyclic aromatic hydrocarbon \citep[PAH; see][for reviews]{Tielens2008, Li2020} flux relative to the dust continuum level, as demonstrated in various studies \citep[e.g.,][]{Madden2006, Engelbracht2005, Wu2006, Gordon2008}, particularly when the metallicity drops to very low levels below 12+log(O/H)$\sim$8 \citep{Draine2007b, Engelbracht2008}.

While there is evidence that PAHs are affected by the metal content and the radiation fields of the ISM, the precise dependence of PAH emission on metallicity, particularly in galaxies of intermediate metallicities, and the role of the interstellar radiation field on PAH emission in such environments warrant further exploration. Several hypotheses have been proposed to address the deficit of PAHs in low-metallicity environments: (i) from a destruction viewpoint, low dust shielding of the ISM in metal-poor environments may allow UV photons to penetrate and dissociate PAHs \citep{Madden2006, Wu2006, Hunt2010, Cormier2015}; (ii) from a chemical formation standpoint, inhibited grain growth in low-metallicity environments may result from a shortage of available carbon atoms essential for building up PAH molecules \citep{Sandstrom2012, Whitcomb2024}; and (iii) \citet{Galliano2008} suggested that AGB stars delay their injection of carbon dust into the ISM, and since these stars are the main source of PAHs \citep{Latter1991}, their absence in young clusters naturally leads to PAH scarcity.

To test these hypotheses, high spatial resolution MIR observations are essential for tracking PAH emission in low-metallicity dwarf galaxies, particularly since observations from Local Group dwarfs reveal that PAH emission concentrates in small clumps, unlike in high-metallicity galaxies where PAHs are pervasively distributed throughout the ISM \citep{Chown2025_WLM}. While Spitzer has provided valuable insights, its spectral mapping capabilities were limited to a handful of low-metallicity targets \citep[e.g.,][]{Haynes2010, Sandstrom2012, Whitcomb2024}. In contrast, JWST’s unparalleled spatial and spectral resolving power has significantly advanced our understanding of stellar populations, as well as the dust and gas properties in metal-poor environments. For instance, previously undetectable populations of dusty evolved stars and young stellar objects have been identified, providing crucial insights into star formation in low-metallicity environments \citep{Hirschauer2024, Lenkic2024, Nally2024}. In the low-mass, isolated galaxy Leo P ($\sim$3\% Z$_{\odot}$), precise photometric detection of stars below the oldest main-sequence turnoff has enabled an accurate reconstruction of its star formation history \citep{McQuinn2024}, and the detection of S(1) 17.03\um\ micron transition of molecular hydrogen in this system has established a lower limit on the metallicity required for molecular hydrogen formation and subsequent star formation \citep{Telford2024}. Moreover, JWST/MIRI has enabled the detection of [Ne V] in the BCD SBS0335-052E ($\sim$5\% Z$_{\odot}$) and I~Zw~18 ($\sim$3\% Z$_{\odot}$), revealing the highly ionized nature in low-metallicity environments \citep{Mingozzi2025, Hunt2025}.

In this paper, we use JWST to investigate the resolved dust properties of \target, a nearby blue compact dwarf (BCD) galaxy located at a distance of 10 Mpc with a metallicity of 12+log(O/H) = 8.1, corresponding to Z = 25\% Z$_\odot$ \citep{Guseva2000}. As one of the earliest studied low-metallicity galaxies \citep{Sargent1970, Searle1972, Bergeron1977}, \target\ exhibits characteristics similar to those of highly ionized \HII\ regions. The galaxy hosts two super-star clusters (SSC-N and SSC-S), with SSC-N, the dominant power source, spanning approximately 15 pc in physical size. This northern cluster hosts intense star formation, with a star formation rate surface density of $1600~\mathrm{M}_\odot\,\mathrm{yr}^{-1}\,\mathrm{kpc}^{-2}$, several times higher than that of 30~Doradus in the Large Magellanic Cloud ($380~\mathrm{M}_\odot\,\mathrm{yr}^{-1}\,\mathrm{kpc}^{-2}$) \citep{Vanzi2008}. Evidence of this vigorous activity is further reflected in \target's exceptionally high \NeIII/\NeII\ ratio, which makes it the highest known for a pure star-forming environment \citep{Hao2009}. This high excitation condition is also evident in HST COS UV spectroscopy \citep{Leitherer2018}, which shows emission line ratios similar to those seen in Green Pea galaxies. Based on SED fitting, dust in \target\ appears to be relatively hot, peaking at $\sim$60\um\ \citep{Galliano2005}. Observations of the young central cluster ($<$5~Myr; \citealt{Kepley2014}) reveal a relatively low gas-to-dust ratio suggestive of a localized dust formation event, as indicated by ALMA observations \citep{Consiglio2016}. Together, these traits make \target\ an exceptional case for studying dust formation and its link to recent star formation in the context of low abundance and high radiation intensity. While PAH emission has been detected in this galaxy via AKARI and Spitzer spectroscopy \citep{Wu2006, Lai2020}, the spatial distribution of the PAH emission was unknown until now.

This paper is organized as follows. \S2 describes the observations and data reduction processes, while \S3 details our analysis. In \S4, we present the resolved PAH properties of \target\ and insights from variations in PAH ratios across different ISM conditions. \S5 proposes a hypothesis on PAH size changes in low-metallicity environments based on these findings, and \S6 offers a summary. For this study, we adopt a redshift for \target\ of $z$=0.00248\footnote{NASA/IPAC Extragalactic Database (NED)}, corresponding to a projected physical scale of 50~pc/arcsec.

\begin{figure*}
\includegraphics[width=1\textwidth]{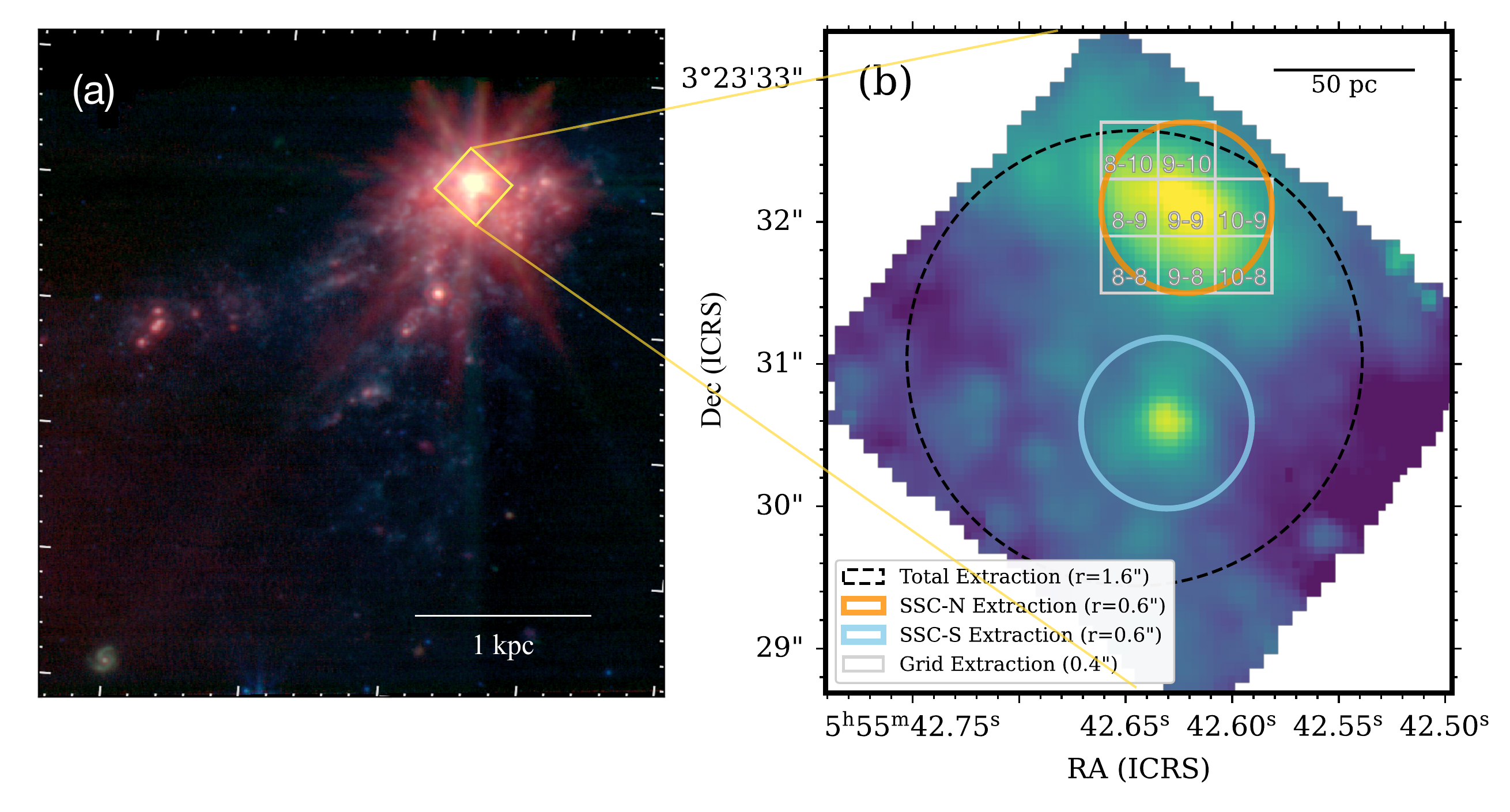}
    \caption{(a) Combined MIRI F770W (blue), F1130W (green), and F2100W (red) images of \target. The originally core-saturated MIRI images were recovered using methods detailed in \S\ref{sec:analysis}. The box indicates the NIRSpec footprint. (b) Spectral extraction regions used throughout this paper. The map is generated by stacking the cube between 3.25---3.34\um. The dashed black circle shows the aperture for total extraction, while orange and blue solid circles represent areas used to extract the SSC-N and SSC-S PAH spectra, respectively, as shown in Figure~\ref{fig:pah33}. Our grid extraction, centered on the northern SSC, yielded eight complete NIRSpec+MIRI/MRS spectra spanning 2.9---24.5\um\ used for resolving PAH properties in \target, with corresponding cell numbers indicated.
}
    \label{fig:IIZw40_img}    
\end{figure*}

\begin{figure*}
\centering
\includegraphics[width=1.\textwidth]{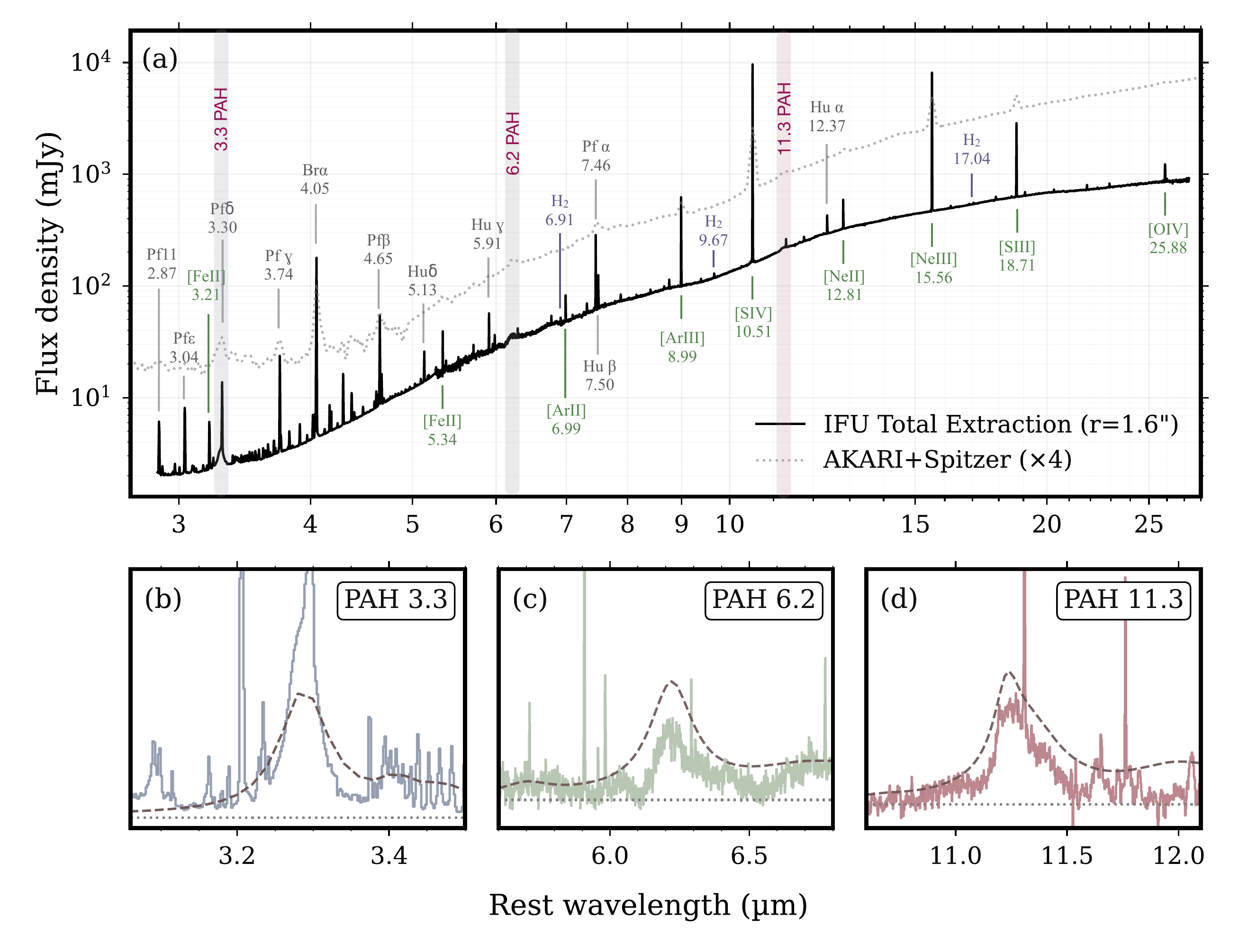}
    \caption{(a) Combined NIRSpec and MIRI IFU spectrum of \target, extracted from a 1\farcs6 radius aperture (dashed black circle in Figure~\ref{fig:IIZw40_img}(b)), showing strong fine structure and recombination lines along with weak \molH\ and PAH features. The three detected PAH bands are highlighted, with other key diagnostics labeled. For comparison, the lower resolution (R$\sim$120) AKARI+Spitzer spectrum is shown as a dotted line, scaled by 4$\times$ for clarity. (b)---(d) Continuum-subtracted zoom-ins of the PAH bands at 3.3, 6.2, and 11.3\um, with dashed profiles representing the 1C PAH template for star-forming galaxies from \citet{Lai2020}, normalized to the 3.3\um\ band strength. In \target, the weaker 6 and 11\um\ bands relative to the template indicate a shift in PAH power toward the 3.3\um\ feature.
}
    \label{fig:tot_spec}    
\end{figure*}

\section{Observations and data reduction}
\label{sec:observation}

The GO program 2511 (PI: T. Lai) includes both the \JWST\ Near-Infrared and Mid-Infrared integral-field spectroscopy taken with NIRSpec integral-field units (IFU) \citep{Boker2022} on Sep 29, 2023 and the Mid-InfraRed Instrument \citep[MIRI:][]{Rieke2015, Labiano2021} on Sep 27, 2023. The data are available at the Mikulski Archive for Space Telescopes (MAST) at DOI: \href{http://dx.doi.org/10.17909/fz5p-qx55}{10.17909/fz5p-qx55}. For the NIRSpec IFU observation, a set of medium-resolution gratings (with a nominal resolving power of R$\sim$1000) was used --- G235M/F170LP and G395M/F290LP, covering the wavelength of 1.66--5.10\um. The science exposure time was 934 seconds, and a 4-pt dither pattern was used to sample the two super-star clusters (SSCs) in \target. For the MRS observations, a set of four IFUs (channels 1–4) was used, covering a full range of 4.9–28.1\um\ with three band settings, SHORT (A), MEDIUM (B), and LONG (C) in each channel, with the science exposure time of 278, 266, and 222 seconds, respectively, and a 4-pt dither pattern.

While taking the MIRI MRS background, simultaneous MIRI images were obtained using the F770W, F1130W, and F2100W filters. These filters were originally selected to capture the two bright PAH emission bands at 7.7 and 11.3\um, while the longer wavelength filter traces the dust continuum at 21\um. 

In this study, we focus on the observations with the NIRSpec G395M/F290LP grating together with Channel 1-4 of MIRI MRS IFUs. The uncalibrated science observations were downloaded through the MAST Portal. For NIRSpec observations, the data reduction process was done using the \JWST\ Science Calibration Pipeline version 1.12.5 \citep{Bushouse2022}, with a reference file of \texttt{jwst\_1177.pmap}. We followed the standard process of the first two stages of the pipeline, including \texttt{Detector1} and \texttt{Spec2}. In \texttt{Spec2}, ``leakcal" was performed in the observation to mitigate the contamination due to a failed open micro-shutter assembly (MSA). For the \texttt{Spec3} stage, which handles the cube build process, at the time of reduction, we substituted it with a routine built in the \texttt{Q3D} software \citep{Rupke2014, Rupke2021} that uses the Python package \texttt{reproject} to construct the cube with a finer spaxel size of 0\farcs05 that properly samples the PSF as opposed to 0\farcs1 from the \texttt{Spec3} stage in the pipeline that is PSF undersampled \citep{Vayner2023}. For the MIRI MRS observation, we followed the data reduction processes based on the Pipeline version 1.15.0, using a reference file of \texttt{jwst\_1276.pmap}. All three stages of the pipeline processing, including \texttt{Detector1}, \texttt{Spec2}, and \texttt{Spec3}, were applied. A change and an additional step in \texttt{Spec2} were conducted --- (i) \textit{residual\_fringe} step is on, and (ii) to correct for the over-subtraction of the continuum near the \ArII\ line at 8.99\um\ caused by the cruciform artifact, an alternative reference file (provided by STScI) for the \textit{straylight} step is adopted. 

\section{Analysis \& Method}
\label{sec:analysis}
The simultaneous MIRI wide band imaging of the F770W, F1130W, and F2100W was taken in parallel with the MIRI IFU observation. Due to the brightness and compactness of the source, these images were saturated at the first frame of the ramp near the northern SSC. Nonetheless, deep MIR images of \target\ can be obtained, featuring the stellar streams that extend $\sim$2~kpc away from the nucleus (see Figure~\ref{fig:IIZw40_img}(a)), which has been suggested as evidence that \target\ underwent a recent merger \citep{vanZee1998, Kepley2014, Beilis2024}. To recover the saturated pixels near the northern SSC, we replaced their values using synthetic photometry derived from matched spectra extracted from the IFU cube, reprojected onto the MIRI images. The recovered 3-color image is presented in Figure~\ref{fig:IIZw40_img}(a), with F770W in blue, F1130W in green, and F1500W in red. 

Combining the NIRSpec and MIRI IFU observations provides a comprehensive analysis of PAH bands across a wide range from 3 to 17\um. While integrating both instruments offers a holistic view of all the PAH emission, the NIRSpec IFU alone provides the advantage of the highest spatial resolution, specifically for probing the smallest PAH grains that emit at 3.3\um. In the first part of this paper (\S~\ref{sec:pah33_variation}), we leverage the 0\farcs05 pixel size of the NIRSpec observations to precisely localize regions within the BCD where small grains can survive and examine the interplay between PAH emission and ionizing gas in a metal-poor environment. The second part of the paper (\S~\ref{sec:pah_ratio}) focuses on the joint NIRSpec and MIRI IFU dataset.  These data sets were combined by smoothing the cubes to a common PSF at 11.5\um, enabling a proper comparison of the 3.3\um\ PAH emission with its longer-wavelength counterparts.

For the NIRSpec cube, we focus on characterizing the aromatic feature at 3.3\um\ and the aliphatic feature at 3.4\um. To achieve this, we perform a detailed spectral decomposition of this wavelength range using the latest version of the \textit{Continuum And Feature Extraction} \texttt{(CAFE)} software \citep{Diaz-Santos2025}. CAFE was originally developed by \citet{Marshall2007} for \emph{Spitzer}/IRS and has been recently updated for \emph{JWST}, incorporating enhanced functionalities for spectral decomposition and the extraction of IFU cubes. The spectrum from each NIRSpec spaxel is fitted with \texttt{CAFE}, covering a wavelength range from 2.87 to 5.27\um, which includes key components such as the 3.3\um\ PAH band, the Pf$\delta$ line, the 3.4\um\ aliphatic feature, the 3.47\um\ plateau feature, and the Br$\alpha$ line. 

For the combined NIRSpec and MIRI cube, we follow the method detailed in \citet{Lai2022}, which smoothed the NIRSpec G395M/F290LP cube and MIRI Channels~1 and 2 by convolving them with wavelength-dependent Gaussian convolution kernels to match with the FWHM measured at 11.5\um\ (FWHM$\sim$0.46\arcsec). Next, we use \texttt{(CAFE)} to perform a 19$\times$19 grid extraction on both smoothed cubes with each cell having a width of 0\farcs4 ($\sim$20 pc), and the grid was set to center at the northern SSC (5:55:42.6324, +3:23:32.000 ICRS). For spectral stitching, CH3 LONG is used as the reference channel, with all spectral segments combined multiplicatively using typical scaling factors of $\sim$2\% for MIRI sub-channels and $\sim$10\% between NIRSpec and MIRI. For the extracted 1D MIRI spectra, an additional spectral defringing procedure was applied using the routine provided by STScI (Kavanagh, P., et al., in prep).

\begin{splitdeluxetable*}{cccccccBcccc}
\tabletypesize{\footnotesize}
\tablewidth{0pt}
\tablecaption{Attenuation-corrected PAH Band and Neon Line Fluxes in \target\ from the Total and Grid Extraction}\label{tab:flux}
\tablehead{
\colhead{Region} & 
\colhead{PAH 3.3\um} & 
\colhead{Aliphatic 3.4\um} & 
\colhead{PAH 6.2\um} & 
\colhead{PAH 11.3\um} &
\colhead{$\Sigma$PAH$_\mathrm{IIZw40}$} & 
\colhead{$\Sigma$PAH$_\mathrm{UL}$} & 
\colhead{Region} & 
\colhead{[NeII] 12.81\um} & 
\colhead{[NeIII] 15.56\um} \\ [-0.2cm]
\colhead{} & 
\colhead{[$\times$10$^{-18}$]} & 
\colhead{[$\times$10$^{-19}$]} & 
\colhead{[$\times$10$^{-18}$]} & 
\colhead{[$\times$10$^{-18}$]} & 
\colhead{[$\times$10$^{-18}$]} & 
\colhead{[$\times$10$^{-18}$]} & 
\colhead{} & 
\colhead{[$\times$10$^{-18}$]} & 
\colhead{[$\times$10$^{-18}$]} \\ [-0.4cm]
}
\startdata
\shortstack{Total \\ Extraction} &  22.06 $\pm$ 2.59 & 21.06 $\pm$ 12.12 & 45.00 $\pm$ 5.52 & 78.32 $\pm$ 10.87 & 145.39 $\pm$ 12.46 & 408.55 $\pm$ 35.02 & 
\shortstack{Total \\ Extraction} & 27.31 $\pm$ 0.37 & 794.05 $\pm$ 42.89\\
\shortstack{SSC-N} & 7.06 $\pm$ 0.67 & 7.31 $\pm$ 3.56 & 26.47 $\pm$ 6.89 & 38.98 $\pm$ 5.91 & 72.51 $\pm$ 9.11 & 203.77 $\pm$ 25.59 & 
\shortstack{SSC-N} & 10.47 $\pm$ 0.57 & 372.91 $\pm$ 27.09\\
\shortstack{SSC-S} & 2.69 $\pm$ 0.25 & 2.16 $\pm$ 1.41 & 8.42 $\pm$ 2.18 & 4.98 $\pm$ 1.45 & 16.09 $\pm$ 2.63 & 42.22 $\pm$ 7.40 & 
\shortstack{SSC-S} & 2.32 $\pm$ 0.06 & 62.11 $\pm$ 2.62\\
\hline
8-8  & 0.32 $\pm$ 0.05 & 0.31 $\pm$ 0.12 & 1.02 $\pm$ 0.24 & 2.46 $\pm$ 0.62 & 3.80 $\pm$ 0.67   & 10.68 $\pm$ 1.88 & 8-8  & 0.84 $\pm$ 0.03 & 29.30 $\pm$ 1.00 \\
8-9  & 0.99 $\pm$ 0.10 & 1.15 $\pm$ 0.55 & 2.93 $\pm$ 0.84 & 5.05 $\pm$ 1.09 & 8.97 $\pm$ 1.38   & 25.21 $\pm$ 3.88 & 8-9  & 1.33 $\pm$ 0.04 & 41.40 $\pm$ 1.38 \\
8-10 & 2.12 $\pm$ 0.10 & 1.97 $\pm$ 0.53 & 5.09 $\pm$ 0.61 & 5.06 $\pm$ 0.61 & 12.27 $\pm$ 0.87 & 34.48 $\pm$ 2.44  & 8-10 & 1.28 $\pm$ 0.06 & 28.30 $\pm$ 1.33 \\
9-8  & 0.59 $\pm$ 0.06 & 0.59 $\pm$ 0.34 & 2.92 $\pm$ 0.35 & 2.80 $\pm$ 0.93 & 6.31 $\pm$ 1.00   & 17.73 $\pm$ 2.81 & 9-8  & 1.25 $\pm$ 0.06 & 42.40 $\pm$ 2.15 \\
9-9  & 1.13 $\pm$ 0.13 & 1.30 $\pm$ 0.74 & 3.26 $\pm$ 1.00 & 6.92 $\pm$ 1.37 & 11.31 $\pm$ 1.70  & 31.78 $\pm$ 4.78 & 9-9  & 1.72 $\pm$ 0.06 & 54.40 $\pm$ 2.51 \\
9-10 & 1.46 $\pm$ 0.10 & 1.48 $\pm$ 0.53 & 3.36 $\pm$ 0.25 & 5.49 $\pm$ 1.27 & 10.31 $\pm$ 1.30 & 28.97 $\pm$ 3.65  & 9-10 & 1.21 $\pm$ 0.04 & 31.50 $\pm$ 1.37 \\
10-8 & 1.23 $\pm$ 0.11 & 1.22 $\pm$ 0.60 & 5.17 $\pm$ 0.66 & 4.83 $\pm$ 0.80 & 11.23 $\pm$ 1.04 & 31.56 $\pm$ 2.92  & 10-8 & 1.40 $\pm$ 0.05 & 39.80 $\pm$ 1.80 \\
10-9 & 1.06 $\pm$ 0.10 & 1.07 $\pm$ 0.24 & 4.38 $\pm$ 0.82 & 5.06 $\pm$ 1.01 & 10.50 $\pm$ 1.31 & 29.50 $\pm$ 3.68  & 10-9 & 1.55 $\pm$ 0.06 & 44.90 $\pm$ 1.63 \\
\enddata
\vspace{0.1cm}
\tablecomments{All the flux units are in W/m$^{2}$. $\Sigma$PAH$_\mathrm{IIZw40}$ is the direct sum of the three PAH features at 3.3, 6.2, and 11.3\um. This value is treated as a lower limit, while $\Sigma$PAH$_\mathrm{UL}$ is the upper limit when accounting for other non-detected PAHs}. See Appendix~\ref{sec:no_pah} for the estimate of $\Sigma$PAH and its upper limit ($\Sigma$PAH$_\mathrm{UL}$).
\end{splitdeluxetable*}

Figure~\ref{fig:IIZw40_img}(b) shows the NIRSpec IFU footprint centering at the two main SSCs in \target. The map is created by stacking the frames between 3.25 and 3.34\um. Overlaid are the apertures for which the spectra in this paper are extracted. The dashed line aperture, centered at (5:55:42.6474, +3:23:31.020 ICRS) with a radius of 1\farcs6, represents the total extraction, with its spectrum presented in Figure~\ref{fig:tot_spec}. Two smaller apertures (each with a radius of 0\farcs6), colored orange and blue, target the regions of SSC-N and SSC-S, respectively; their extracted spectra appear in Figure~\ref{fig:pah33} and are further explored in \S\ref{sec:pah33_variation}.

For the grid extraction, we focus on the eight spectra extracted from cells near the northern SSC, as marked in Figure~\ref{fig:IIZw40_img}(b), selected for their high spectral quality across both NIRSpec and MIRI wavelength ranges. Spectra from regions farther from the clusters, however, suffer from low S/N, especially at the blue end of MIRI (CH1), where MIRI’s lower sensitivity compared to NIRSpec, the source’s red MIR spectral SED, and its compact size pose challenges to achieving adequate spectral quality. The need for high S/N across all wavelengths is primarily driven by our scientific goal to accurately measure all available PAH bands from 3---17\um. With the combined NIRSpec and MIRI observations, we find in \target\ the only detected PAHs are the 3.3, 6.2, and 11.3\um\ bands, where 11.3\um\ is a PAH complex that includes both the 11.23 and 11.33\um\ bands. 

Given that the spectra are continuum-dominated, simultaneous spectroscopic fitting with a wide wavelength range across NIRSpec and MIRI is challenging. Therefore, to precisely measure the PAH fluxes, we fit each band locally with carefully defined spectral ranges, with [3.0, 3.6], [5.6, 7.2], and [10.7, 12.4] for PAH 3.3, 6.2, and 11.3\um\ bands, respectively. These fluxes are subsequently corrected for attenuation, assuming a mixed geometry based on the opacity profile fitted by CAFE using the full NIRSpec+MIRI spectrum. This attenuation correction is minimal, with the correction factor for the 11.3\um\ PAH emission, the band most susceptible to attenuation \citep{Lai2024}, reaching only the 5\% level. Neon lines are corrected in the same manner as the PAHs. Table~\ref{tab:flux} lists all attenuation-corrected measurements of PAH and neon line fluxes. 

Also included in Table~\ref{tab:flux} are the fluxes of total PAH power, with their corresponding lower and upper limits ($\Sigma$PAH$_\mathrm{IIZw40}$ and $\Sigma$PAH$_\mathrm{UL}$). $\Sigma$PAH$_\mathrm{IIZw40}$ is the direct sum of the robustly detected PAH bands at 3.3, 6.2, and 11.3\um. This value is considered a \emph{lower limit} in our estimation, given that contributions from other typically prominent PAH features (e.g., 7.7, 8.6, and 17\um\ PAHs) are not included. The effects of continuum dilution on the detectability of different PAH features depend on the strength of the underlying continuum and the power-to-width ratio of the feature.  Broad features (PAH 7.7 and 17) in particular fall beneath the detection floor more readily than narrow features (e.g., PAH 3.3).  While certain broad features such as PAH 7.7, 8.6, and 17 may be present below our detection threshold, we can directly estimate this missing power using known templates. This estimation, detailed in Appendix~\ref{sec:no_pah}, yields an upper limit for the total PAH flux $\Sigma$PAH$_\mathrm{UL}$ being 2.81$\times$ of $\Sigma$PAH$_\mathrm{IIZw40}$.

\section{Results}
\label{sec:Results}
The Field of View (FoV) of the dithered JWST NIRSpec and MIRI IFU observations provides complete coverage of the two main SSCs (SSC-N and SSC-S) in \target, allowing for high spatial and spectral resolution in resolving dust and star-formation properties on a scale of a few tens of pc. As mentioned in \S\ref{sec:analysis}, the total IFU spectrum was extracted with a radius of 1\farcs6, as indicated by the dashed line in Figure~\ref{fig:IIZw40_img}(b), encompassing the largest aperture size that fits within the FoV.

The extracted spectrum, presented in Figure~\ref{fig:tot_spec}(a), is dominated by hot dust continuum underlying strong atomic emission lines, with relatively faint PAH emission appearing atop the continuum. Among these PAH features, the 3.3\um\ band exhibits the greatest contrast to the continuum. Panels (b)–(d) show the continuum-subtracted profiles of the 3.3, 6.2, and 11.3\um\ bands, with dotted horizontal lines indicating the zero baseline. The dashed lines represent the modified 1C template from \citet{Lai2020}, featuring the continuum-subtracted PAH and aliphatic emission, scaled to match the 3.3\um\ PAH flux in \target. This template, derived from low spectral resolution AKARI and Spitzer spectra, exhibits a broader width primarily due to its lower spectral resolution of R$\sim$120. Both the 6.2 and 11.3\um\ PAH profiles appear weaker than the reference starburst template, suggesting a PAH power shift toward PAH 3.3. Within the \target\ spectrum, only the 3.3, 6.2, and 11.3\um\ PAH features are robustly detected. 

A suite of atomic and rotational \molH\ lines can also be seen. Notably, we detected the \OIV\ line at 25.89\um, which typically traces AGN activity \citep{Genzel2000, Peeters2004a, LaMassa2010} but can also appear in environments with intense star formation \citep{Hunt2006}, particularly in the presence of metal-poor Wolf-Rayet stars \citep{Tarantino2024}. 

The AKARI+Spitzer spectrum of \target\ from \citet{Lai2020} is shown in Figure~\ref{fig:tot_spec}(a). While the overall spectral shapes are in good agreement, JWST's higher spectral resolution allows for the detection of additional emission lines. In particular, the Pf$\delta$ line, which overlaps with the 3.3\um\ PAH band, can now be clearly separated from the PAH feature --- a result that was not possible with the lower resolution AKARI spectrum. 

In \citet{Lai2020}, the blending of 3.3\um\ PAH and Pf$\delta$ was accounted for when estimating the PAH 3.3 flux by subtracting the possible contribution from the Pf$\delta$ line, scaled from Br$\alpha$ assuming Case B conditions. With JWST, we find the 3.3\um\ PAH flux measured in the total extraction is $2.21\times10^{-17}$ W/m$^{2}$, approximately 25\% of the 3.3\um\ flux estimated in the AKARI+Spitzer spectrum. This discrepancy is likely due to two reasons. First, the AKARI+Spitzer spectrum essentially probes a region that is more than 10 times larger than our IFU-extracted area\footnote{The AKARI+Spitzer spectrum is drawn from the IDEOS sample which used the Spitzer IRS LL2 (with a slit width of 10\arcsec) as a reference to stitch the shorter wavelength segments, and the AKARI spectrum was then scaled to match with the IRS spectrum \citep{Lai2020}.}. Second, with the AKARI spectral resolution of R$\sim$120, multiple emission lines were blended. Other bright emission lines, in addition to Pf$\delta$ (e.g. \FeII\ at 3.21\um), lie close to the PAH 3.3 band and were not fully subtracted in the low-resolution AKARI measurement, resulting in an overestimate of the 3.3\um\ PAH flux.

\subsection{Spectral Variation across \target}
\label{sec:pah33_variation}
\begin{figure}
\includegraphics[width=1\columnwidth]{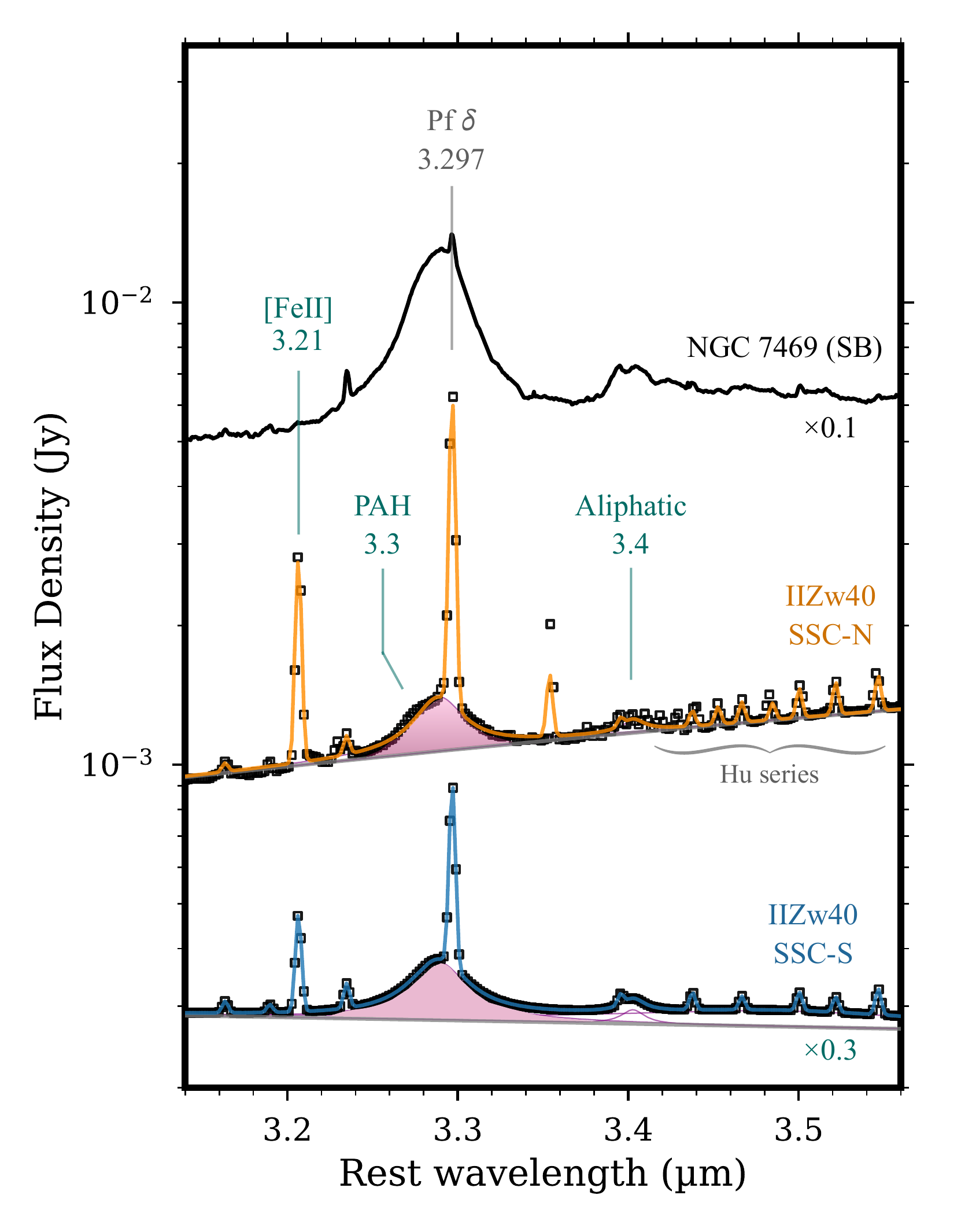}
    \caption{The 3.3\um\ PAH emission in \target. The two \target\ spectra, extracted from the apertures outlined in Figure~\ref{fig:IIZw40_img}(b), represent SSC-N (orange) and SSC-S (blue) regions. The CAFE fits to the 3.3\um\ PAH feature and nearby lines and continuum in \target\ are shown, with the PAH 3.3 bands shown in shaded purple. The 3.47\um\ plateau feature can be seen in SSC-S. Also shown is the starburst template spectrum from NGC~7469 in \citet{Lai2023}. Notably, \target\ displays a significantly different relative flux ratio between the 3.3\um\ PAH band and the $\Pfdelta$ line compared to typical star-forming galaxies, with the $\Pfdelta$ emission being exceptionally strong. For better visibility, the spectra of SSC-S and the starburst ring in NGC~7469 are scaled as indicated.
    }
    \label{fig:pah33}    
\end{figure}

With JWST's enhanced spatial/spectral resolving power, we see that PAH emission varies significantly across the face of \target. For illustration, Figure~\ref{fig:pah33} shows the 3.3\um\ PAH profiles from the SSC-N and SSC-S regions extracted from the orange and blue circles, respectively, indicated in Figure~\ref{fig:IIZw40_img}(b). Also included is the star-forming ring spectrum in NGC~7469 \citep{Lai2023}, which represents the typical 3.3\um\ profile in galaxies with active star formation. Compared to the NGC~7469 spectrum, \target\ spectra exhibit prominent Pf$\delta$ emission relative to the PAH 3.3 component due to the strong recombination line plus the suppression of the PAH emission. The relative contribution between the 3.3\um\ PAH and the Pf$\delta$ line across the IFU and its photometric implications will be discussed in \S\ref{sec:synphot}.

\begin{figure*} 	\includegraphics[width=1\textwidth]{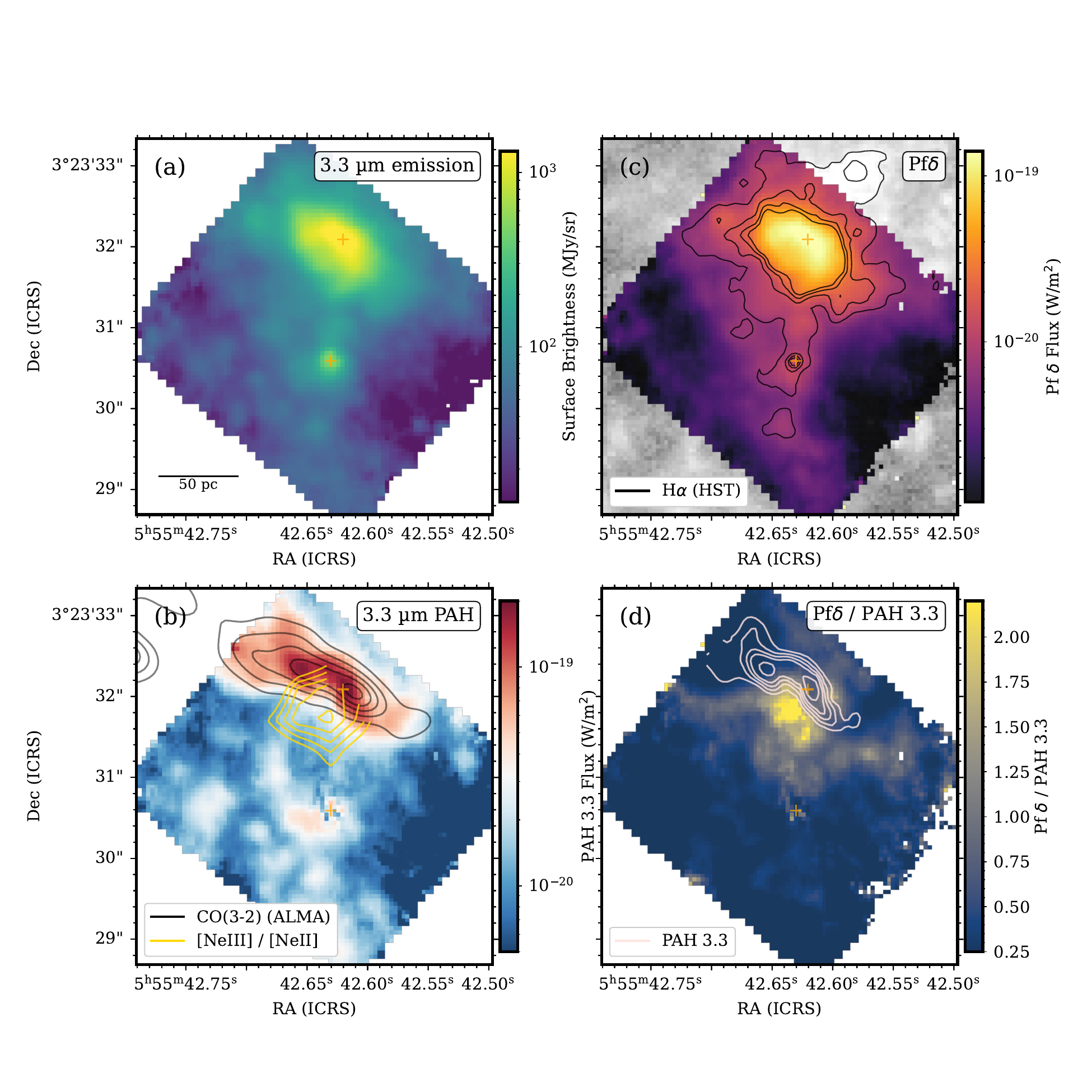}
    \caption{High spatial resolution maps from NIRSpec, with a pixel size of 0\farcs05 and a typical PSF of $\sim$0\farcs19 at 3.3\um. (a) A map of the 3.3\um\ emission, highlighting the two main super star clusters (SSCs) in \target, indicated by the ``+" signs. (b) and (c) Spectral decomposition reveals two distinct components: (b) The 3.3\um\ PAH map, uncovering its elongated morphology not visible in (a). Overlaid CO (3–2) contours from ALMA \citep{Consiglio2016} indicate co-spatiality with 3.3\um\ PAH, while yellow contours represent the \NeIII/\NeII\ ratio. The offset between the 3.3\um\ PAH emission and neon ratio is intriguing, resembling blisters found at the edges of \HII\ regions; and (c) The Pf$\delta$ map, shown alongside an H$\alpha$ grayscale map from HST, where the H$\alpha$ contour closely aligns with Pf$\delta$, suggesting a strong correlation. (d) The Pf$\delta$/PAH 3.3 ratio map, overlaid with the PAH 3.3 contour from (b), shows a peak displaced from the 3.3\um\ PAH emission, highlighting the spatial offset between PAH 3.3 and Pf$\delta$.
    }
    \label{fig:IIZw40_maps}
\end{figure*}

Spectral decomposition using CAFE \citep{Diaz-Santos2025}\footnote{https://github.com/GOALS-survey/CAFE} is applied to each individual spaxel (with 0\farcs05 width) within the NIRSpec cube, totaling approximately $\sim$5000 spaxels. Figure~\ref{fig:IIZw40_maps}(a) presents the 3.3\um\ flux map, with crosses marking the peak emission in both SSCs, providing reference points for subsequent maps. Panels (b) and (c) display the resulting spectral fits, the 3.3\um\ PAH and Pf$\delta$ distributions, respectively. These maps reveal distinct morphologies between the emitters: the 3.3\um\ PAH emission is concentrated at SSC-N with an elongated structure, whereas the Pf$\delta$ emission appears more rounded and diffuse. The bar-like distribution of the 3.3\um\ PAH emission emerges only after isolating the PAH component through spectral decomposition. This PAH emission correlates well with the CO (3-2) emission from ALMA (ID: 2012.1.00984.S; PI: A. Kepley), indicating the co-spatiality of PAHs and CO \citep[e.g.,][]{Shivaei2024, Chown2025_PAH_CO, Shim2025}. This co-spatiality suggests that shielding from the molecular gas is likely conducive to the survival of these small PAHs, protecting them from destruction by the radiation field. The contours of the neon ratios (\NeIII/\NeII) from the MIRI observation are also included in Panel (b). The morphological correspondence between the bend in the 3.3\um\ PAH emission map and the neon emission is reminiscent of blisters at the edge of \HII\ regions. The grayscale background and the contours in Panel~(c) are from H$\alpha$ emission observed by HST (ID: 9739; PI: R. Chandar). We see the $\Pfdelta$ emission agrees well with the H$\alpha$ emission. For Panel~(d), the map shows the ratios of $\Pfdelta$ to 3.3\um\ PAH, and the contours are from PAH 3.3 (Panel (b)). As opposed to main sequence star-forming galaxies, which have relatively weak $\Pfdelta$ emission relative to 3.3\um\ PAH emission (e.g., the ratio of $\Pfdelta$ to PAH 3.3 in the star-forming ring of NGC~7469 is $\sim$0.01), \target\ exhibits a high $\Pfdelta$/PAH 3.3 ratio, reaching $\sim$2 in regions close to the illuminated stars, coinciding with the peak in the \NeIII/\NeII\ contour that traces radiation field hardness. 


\begin{figure}
\includegraphics[width=1\columnwidth]{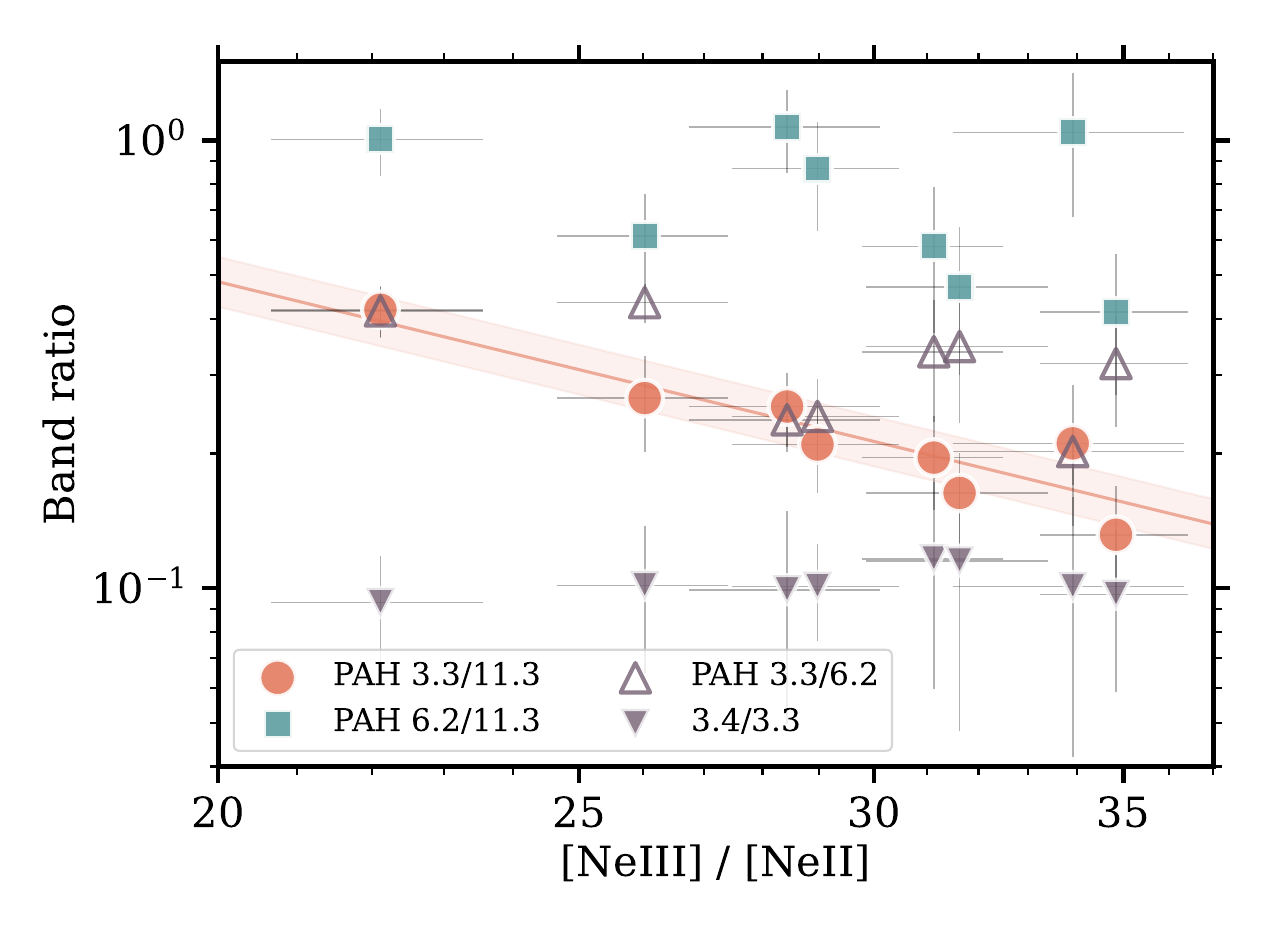}
    \caption{PAH (and aliphatic) band ratios plotted against \NeIII/\NeII, which probes radiation field hardness. Different symbols represent various band ratios as shown in the legend. Only the PAH 3.3/11.3 ratio exhibits a significant correlation (r=-0.92), with a 1-$\sigma$ spread of $\sim$0.06 dex (shaded), suggesting larger average PAH sizes in regions with harder radiation fields. Other band ratios show no significant variations. 
    }
    \label{fig:Ne_pah_ratio}    
\end{figure}

\subsection{Resolving PAH band ratios in \target}
\label{sec:pah_ratio}
In this section, we focus on the results carried out by the combined NIRSpec and MIRI cubes. As pointed out in \S\ref{sec:analysis}, we are able to extract eight independent spectra from the combined cubes that span a wavelength range of 2.87---27.90\um. For all eight spectra, the most prominent PAH band detected is the 3.3\um\ PAH feature, whereas other PAH bands at longer wavelengths are weak and hardly distinguishable from the underlying continuum, which rises steeply to the red. 

We compare the PAH ratios derived from each spectrum with the neon line ratios measured by MIRI/MRS, with the results in Figure~\ref{fig:Ne_pah_ratio}. The four symbols in the figure represent the combined ratios of PAH (and aliphatic) bands relative to the \NeIII/\NeII\ ratio, which is used primarily as a diagnostic of the radiation field hardness, albeit with a secondary dependence on the ionization parameter \citep{Thornley2000, Forster_Schreiber2001}. Here the \NeIII/\NeII\ ratio reaches as high as 35 in cell 8-8, a factor of 1.75 higher than the maximum seen with Spitzer/IRS \citep{Hao2009}, likely due to the higher spatial resolution in the JWST data. 

The PAH ratios are expressed as short-to-long wavelength ratios, where a higher value indicates a smaller PAH size distribution and a lower value corresponds to larger PAH sizes \citep{Schutte1993}. The 3.3/11.3 ratio exhibits a strong correlation (Pearson r=-0.92) to \NeIII/\NeII, while other ratios show no significant variation. The high correlation coefficient between the \NeIII/\NeII\ line flux ratio and 3.3/11.3 suggests that the radiation field is correlated with the PAH sizes in \target, since this PAH ratio is a strong tracer of average PAH size \citep{Croiset2016, Maragkoudakis2020}. While this finding is consistent with previous studies that support the photodestruction of PAHs in metal-poor environments \citep{Madden2006, Wu2006, Gordon2008, Hunt2010}, it is worth noting that no 17\um\ PAH emission, which is typically associated with the largest PAH populations, is detected in \target. 

While the aromatic-to-aliphatic ratio is known to decrease with increasing intensity and hardness of the radiation field \citep[e.g.][]{Joblin1996, Pilleri2015, Lai2023, Peeters2024, Chown2024}, in this case, no significant correlation was observed with the \NeIII/\NeII\ ratio. Further investigation into its cause requires additional observations of aliphatic features in dwarf galaxies. The plateau feature centered at $\sim$3.47\um\ appears only in two cells, 8-10 and 9-10, which have relatively low \NeIII/\NeII\ ratios and are co-spatial with the CO molecular cloud that provides additional shielding. This lack of detection of the broad 3.47\um\ plateau suggests that this feature disappears more readily than the 3.3 and 3.4\um\ bands as radiation hardness increases.

\begin{figure*}
    \centering        \includegraphics[width=0.75\textwidth]{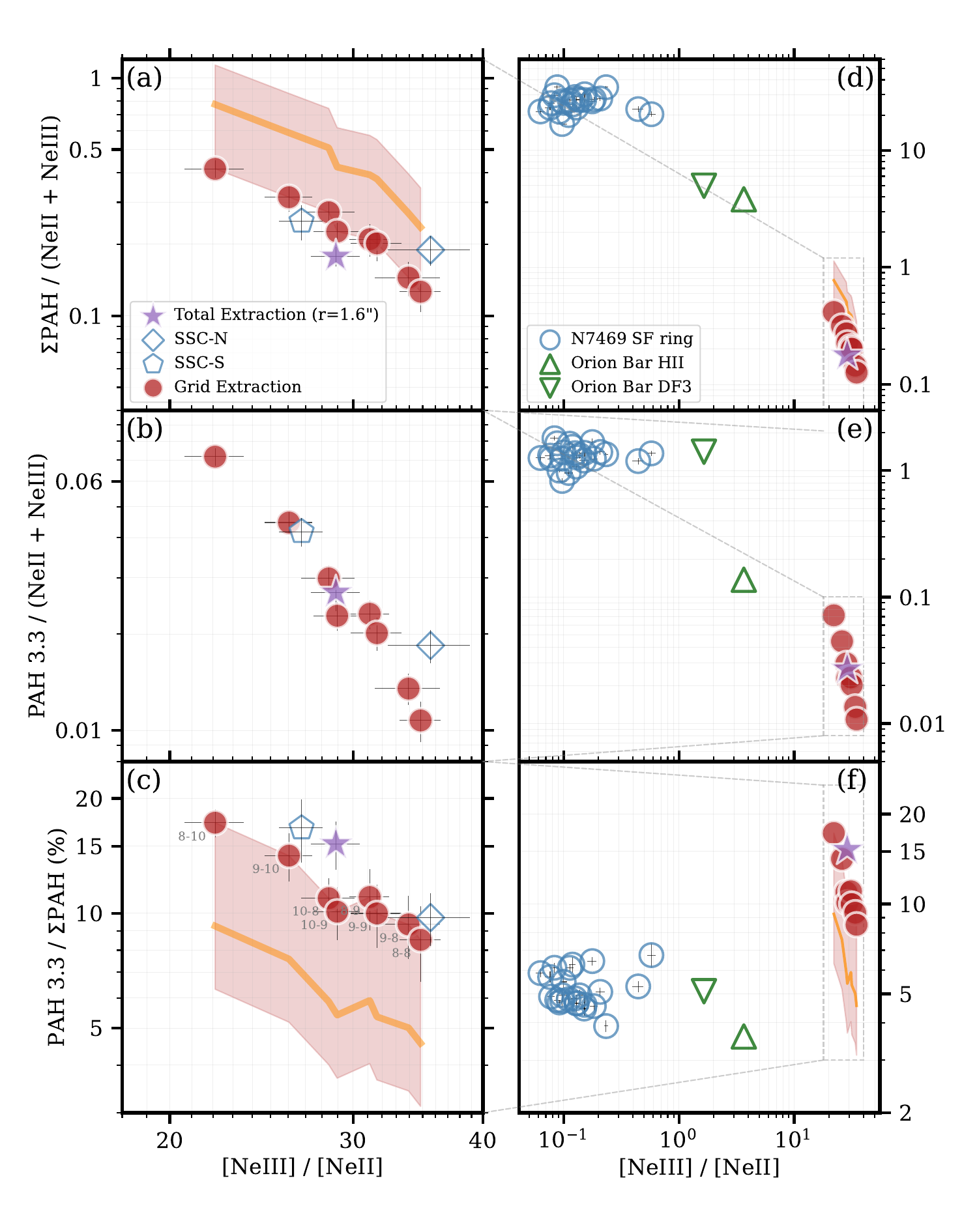} 
    \caption{Left panels display results from \target, while right panels provide context by comparing them to other star-forming regions in NGC~7469 \citep{Lai2023} and photodissociation regions (PDRs) in the Orion Bar. The Orion Bar data include the \HII\ and diffuse front (DF3) from \citet{Chown2024} and \citet{Peeters2024}. Red points denote individual cell results, with the purple star showing the total extraction. Results from both SSC-N and SSC-S are also shown; overall, the different extractions exhibit a consistent trend. A red shaded region illustrates the propagated limit for ratios involving $\Sigma$PAH, with the orange solid line indicating the midpoint of our possible distribution (see Section~\ref{sec:pah_formation_destruction} for more detail). Panels (a)---(b) demonstrated that PAH emission becomes increasingly suppressed relative to ionizing neon emission as radiation fields harden. Panel (c) shows the decrease of the fractional 3.3\um\ PAH power with respect to the neon ratio as the galaxy is resolved. Panel (d) reveals that PAH emission is heavily suppressed in \target, with $\Sigma$PAH-to-neon ratios $\sim$100 times lower than those in NGC~7469's star-forming ring and $\sim$10 times lower than the Orion Bar. However, as shown in panel (e), PAH 3.3 suppression is slightly less severe, decreasing only about 60$\times$ compared to NGC~7469. Panel (f) illustrates how this non-uniform suppression of PAHs and greater retention of 3.3\um\ emission leads to the enhancement of the fractional PAH 3.3 strength. Based on the conservative midpoint distribution, this enhancement can reach $\sim$10\%—at least twice the level observed in reference star-forming regions, underscoring the notable shift of PAH power towards the 3.3\um\ PAH in \target.
    } 
    \label{fig:pah33_totPAH_Ne}
\end{figure*}

\section{Discussion}
The smallest condensed dust grains, while emitting only a tiny share of a galaxy's total dust luminosity, are in many ways the most responsive to their environments.  Thanks to their tiny heat capacities, they are uniquely sensitive to the effects of destruction by photons, elevated radiation intensity and changes in mean photon energy, and may form the building blocks for growing larger grains through coagulation.  Here we discuss how these competing effects may operate in galaxies like \target.

\subsection{Tug-of-war: Competing Mechanisms that Shape PAHs in \target}
\label{sec:tug_of_war}
The influence of metallicity on the life cycle of PAHs has been a subject of investigation for decades, particularly with Spitzer spectroscopy, which provided access to prominent PAH bands longward of 5\um. \citet{Hunt2010} examined 22 blue compact dwarfs (BCDs) and identified a significant deficit---exceeding an order of magnitude---in PAH emission relative to total infrared luminosity in low-metallicity dwarfs compared to metal-rich star-forming galaxies. The authors attributed this deficit of PAHs to their destruction by intense, hard radiation fields. Similar conclusions were drawn by other studies, which identified photodissociation as the dominant mechanism processing PAH grains in low-metallicity environments \citep{Martin-Hernandez2002, Madden2006}. Yet, observations of normal star-forming galaxies in SINGS \citep{Kennicutt2003} noted the suppression of large-grains towards lower metallicity \citep{Smith2007}, and star-forming regions in the Small Magellanic Cloud revealed a shift in PAH size distribution toward smaller grains, challenging the idea that radiation-driven PAH destruction is the dominant effect \citep{Sandstrom2012}. Instead, the authors argued that differences in PAH formation at low metallicity primarily govern PAH size distribution. Mid-infrared spectroscopic maps of nearby galaxies from Spitzer combined with detailed grain modeling have further supported this perspective \citep{Whitcomb2024}. More recently, JWST IFU observations of 30 Doradus reached a similar conclusion, showing that PAHs in this low-metallicity environment tend to be smaller than those in the Orion Bar \citep{Zhang2024}. Indeed, if photo-destruction were the dominant mechanism, the presence of 3.3\um\ emission would be unexpected, as its carriers, which represent the smallest PAH population, are thought to be the most vulnerable to photo-processing. Nevertheless, recurrent fluorescence could be a plausible mechanism to enhance the survival of these small molecules by providing extremely efficient energy dissipation pathways \citep[e.g.,][]{Leger1988, Lai2017, Draine2021}. 

\subsubsection{PAH Formation and Destruction in Low Metallicity Environments}
\label{sec:pah_formation_destruction}
It is now possible to study PAH emission, in particular the 3.3\um\ PAH feature, in low-metallicity galaxies at unprecedented sensitivity with JWST, providing a comprehensive understanding of the size distribution in low-metal systems via the 3.3/11.3 ratio  \citep{Maragkoudakis2020}. Figure~\ref{fig:Ne_pah_ratio} exhibits a 3.3/11.3 PAH ratio ranging from $\sim$0.13---0.42, corresponding to PAH molecules with $\sim$70---150 carbon atoms \citep{Croiset2016, Maragkoudakis2020, Rigopoulou2021}. This ratio reveals a strong correlation between the neon ratio and PAH 3.3/11.3 in \target\, suggesting small PAHs are photo-processed as they approach the bright \HII\ regions. However, this top-down scenario cannot fully explain two critical observations in \target: the absence of the 17\um\ PAH band, which typically traces the largest PAH population, and the unexpectedly strong presence of the 3.3\um\ relative to the longer wavelength PAH bands that trace larger grains. Photodissociation is clearly not the dominant mechanism determining the size distribution of PAHs in \target. If it were, the grains traced by the 3.3\um\ band should be the \emph{first} to disappear, leaving only the largest PAH populations. Other mechanisms must be at play in \target.

Following the analysis presented in \citet{Lai2023}, we study the comparison between the PAH and neon ratios and how they compare to star-forming and photodissociation regions in NGC~7469 \citep{Lai2023} and the Orion Bar \citep{Chown2024, Peeters2024}. As shown in Figure~\ref{fig:pah33_totPAH_Ne}(d), the total PAH emission ($\Sigma$PAH) in \target, when normalized to the \NeII+\NeIII\ flux, is more than an order of magnitude lower than in the NGC~7469 and the Orion Bar, which agrees with the observations in \citet{Hunt2010} where the average of $\Sigma$PAH/TIR in the BCDs is $\sim$0.5\% (it is closer to $\sim$10\% in metal-rich star-forming galaxies). While total PAH emission is low in \target\ compared to high-metallicity starburst galaxies, the 3.3\um\ PAH is less suppressed (see Panel (e)), leading to an elevated fractional 3.3\um\ PAH strength (PAH 3.3/$\Sigma$PAH) as shown in Panel~(f).

In some regions with moderate neon ratios (\NeIII/\NeII$\lesssim$30), the fractional PAH 3.3 strength can reach values as high as $\sim$10\%. These elevated PAH 3.3/$\Sigma$PAH ratios agree well with the predictions of the \emph{inhibited growth model} proposed by \citet{Whitcomb2024}, which showed that fractional 3.3\um\ PAH power can reach a factor of $\sim$3--4 higher than the typical solar metallicity values as indicated by the contours in Figure 4 of \citet{Lai2023}. These projections have been confirmed in new NIRCam measurements of 3.3\um\ emission in M101, which show 3.3/$\Sigma$PAH ratios in the range of 5---10\% (Whitcomb et al., in prep.).

An additional test was conducted to validate the 3---6\% lower limit of PAH 3.3/$\Sigma$PAH, as depicted in Figure~\ref{fig:pah33_totPAH_Ne}(c). By scaling a PAH-metallicity model spectrum ($Z = 0.25\,Z_{\odot}$) from Whitcomb et al. (in prep.) to match the observed PAH 6.2 \um\ flux, we recalculated $\Sigma$PAH to estimate upper limits that account for missing PAH features. The resulting PAH 3.3/$\Sigma$PAH ratios, ranging from 3--7\% across all spatial cells, are in good agreement with the quoted lower limits (lower envelope of the shaded region). These two independent and conservative estimates collectively indicate no relative decrease in PAH 3.3 when compared to star-forming region references.

Taken together, these observations reveal that the PAH size distribution in \target\ shows a marked shift toward smaller molecules compared to typical star-forming regions. Although the fractional strength of the 3.3\um\ PAH to total PAH emission can vary considerably when taking into account the limits, spanning from approximately 3\% to 17\% in this spatially resolved study, there is \emph{no} evidence that small PAH molecules are the first to vanish in metal-poor environments.  On the contrary, these tiniest of grains appear to be robust. Therefore, we conclude that inhibited growth may also play a key role in shaping the PAH size distribution.  The influence of photo-processing of the smallest grain cannot, however, be fully dismissed, as evidenced by the tight 3.3/11.3-to-\NeIII/\NeII\  correlation seen in Figure~\ref{fig:Ne_pah_ratio}, where the ISM is resolved on a 20 pc scale. \target, with its metallicity of approximately 0.25~Z$_{\odot}$, provides a unique opportunity to observe the interplay and competition between the top-down and bottom-up evolution path of PAHs, as galaxies with even lower metallicity might not show the presence of PAHs even with JWST \citep{Telford2024, Mingozzi2025}. This underscores the value of studying PAH properties in galaxies with intermediate metallicities, where these effects are more evident.

\subsubsection{Small-PAH Imposter: PAH with double photon heating}
In addition to the two mechanisms discussed above which can alter the PAH size distribution, an alternative explanation for the disappearance of the 17\um\ PAH feature together with the enhanced 3.3\um\ PAH emission is a power shift from long to short wavelengths in large PAH molecules when exposed to intense radiation fields with $U \gtrsim 300$ \citep{Draine2021}. In such a high-$U$ environment, a large PAH grain can behave like a small-PAH ``impostor" by absorbing an additional UV photon before fully cooling, thereby shifting its emission toward shorter wavelengths. The radiation field intensity in SSC-N is exceptionally high, with $U \sim 76{,}000$, corresponding to a Habing unit of \Gnought$\sim 8.6 \times 10^4$ (see Appendix~\ref{sec:radiation_field} for details). This level of radiation is consistent with the $U$ value at which the 17\um-to-total IR flux ratio reaches a minimum before its rise due to the second peak associated with a bimodal PAH size distribution \citep[see][]{Draine2021}. In other words, a substantial portion of the power that would have been emitted at 17\um\ is instead redistributed to shorter-wavelength emission. The extreme radiation field in SSC-N could therefore account for the absence of the 17\um\ PAH feature and the strong emission at 3.3\um\ in \target.

\begin{figure}
    \centering
        \includegraphics[width=1\columnwidth]{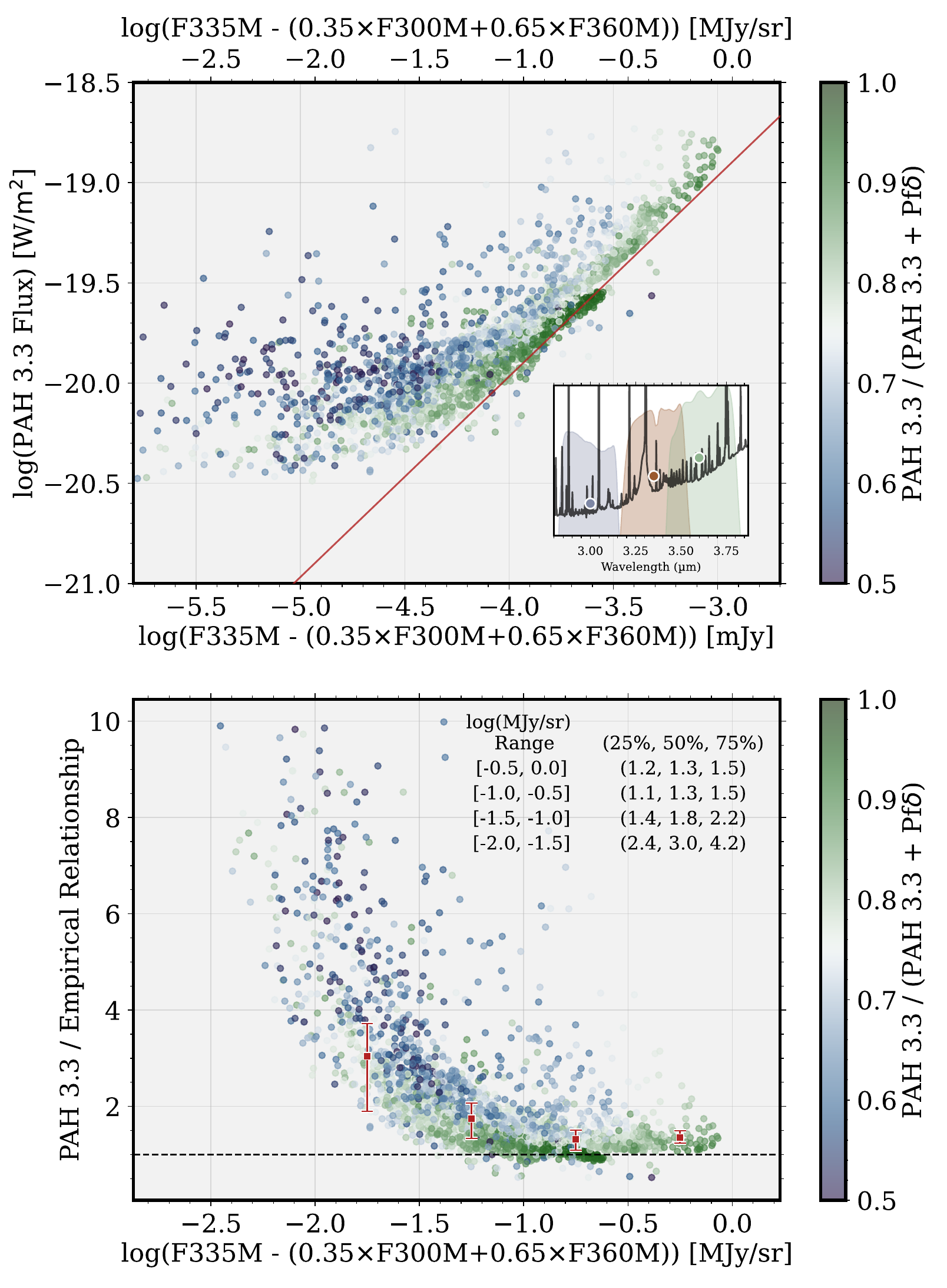} 
    \caption{(\textit{Top panel}) The comparison between the 3.3\um\ PAH flux based on the continuum subtraction method using the three NIRSpec mid-band filters and the result from spectral decomposition. Each point is color-coded by the fractional contribution of the 3.3\um\ PAH to the combined PAH and $\Pfdelta$ emission.  The red line represents the \citet{Lai2020} photometric PAH 3.3 empirical relationship, which aligns well with regions where PAH emission dominates over the $\Pfdelta$ line. The inset demonstrates the relative positions of the filters and the derived synthetic photometry with respect to the spectrum from the total extraction shown in Figure~\ref{fig:tot_spec}. (\textit{Bottom panel}) The division between the spectroscopically measured PAH 3.3 and the empirical relationship from the top panel. The x-axis is expressed in surface brightness. Based on the binning results (in red), the photometric method can recover the ``true" 3.3\um\ PAH flux within 50\% when the continuum-subtracted F335M surface brightness exceeds 0.1 MJy/sr. 
    } 
    \label{fig:pah33_synphot}
\end{figure}

\subsection{Probing 3.3\um\ PAHs Photometrically in Dwarfs}
\label{sec:synphot}
Various methods have been developed to recover the 3.3\um\ PAH photometrically using the NIRCam filters \citep[e.g.,][]{Sandstrom2023, Bolatto2024, Gregg2024}. With our NIRSpec spectroscopy, we compare synthetic photometry derived from three NIRCam filters (F300M, F335M, and F360M) using the \texttt{pyphot} software \citep{Morgan2025} with spectroscopically fitted 3.3\um\ PAH fluxes. This comparison, illustrated in Figure~\ref{fig:pah33_synphot}(\textit{top panel}), incorporates all spaxels within the total extraction region shown in Figure~\ref{fig:IIZw40_img}(b), and evaluates how effectively 3.3\um\ PAH emission can be recovered through photometric methods. Each point represents measurements from a single spaxel with a width of 0\farcs05, with the corresponding surface brightness on the top axis. The empirical formula for probing 3.3\um\ PAH flux as established by \citet{Lai2020} can be written as
\begin{equation}
    \bigg(\frac{f_\mathrm{PAH \ 3.3}}{10^{-17} \ \mathrm{W m^{-2}}}\bigg) = (10.78 \ \pm \ 0.44) \cdot x, 
\end{equation}
where
\begin{equation}
    x = \mathrm{F335M} - (0.35\times\mathrm{F300M} + 0.65\times\mathrm{F360M}) \ \mathrm{[mJy]}.
\end{equation}
This relationship is indicated by the red line, and each point in the figure is color-coded by the ratio of PAH 3.3 / (PAH 3.3 + Pf$\delta$) in each spaxel. Notably, points with relatively higher contributions from the 3.3\um\ PAH align closely with the empirical line. As the $\Pfdelta$ contribution to the F335M filter increases, points progressively deviate from the established relationship. This observation supports our prediction, as the empirical relationship was derived from a set of starburst galaxies that exhibit relatively minimal $\Pfdelta$ emission (similar to the NGC~7469 ring spectrum shown in Figure~\ref{fig:pah33}). In contrast, we find the 5---95\% range of PAH 3.3 / ($\Pfdelta$+PAH 3.3) is 54---89\%, indicating that $\Pfdelta$ can account for approximately 10---50\% of the residual power after continuum subtraction in F335M. Therefore, caution should be exercised when probing 3.3\um\ PAH emission with photometry in dwarfs, particularly in highly ionized environments. One potential approach to mitigate contamination from $\Pfdelta$ is to obtain observations of additional recombination lines, such as Pa$\alpha$ and Br$\alpha$, using the F187N and F405N narrow-band filters, although this may require substantial observing time.

To further understand how accurately the 3.3\um\ PAH flux can be recovered photometrically, we divide the CAFE-measured PAH 3.3 flux by the empirical relationship (Figure~\ref{fig:pah33_synphot} bottom panel) and bin the data according to surface brightness. Most points show ratios larger than unity because the artificial continuum level derived from the F300M and F360M filters is likely overestimated in most cases due to the numerous strong emission lines present within the filters, as shown in the inset of Figure~\ref{fig:pah33_synphot}(top panel). Based on the binning results, we show that the photometric method can recover the 3.3\um\ PAH flux within 50\% of the ``true" value when the continuum-subtracted F335M surface brightness is greater than 0.1 MJy/sr in \target. The precision of using NIRCam filters to probe PAH 3.3 deteriorates drastically (uncertainty increases 2---3$\times$) below the 0.1 MJy/sr threshold. We emphasize that these findings are specific to \target\ and anticipate that in galaxies with even lower metallicity, deviations from the empirical trend will occur at even higher surface brightness thresholds. Fully understanding how this trend varies with metallicity will thus require additional observations.

\section{Summary}
In this paper, we present a study of resolved PAH emission on $\sim$20 pc scales in \target, a BCD with metallicity of 25\% of solar, using JWST NIRSpec and MIRI. In particular, we focus on pinpointing the location of strong PAH emission and the relation between PAH emission and the local radiation field. Our findings can be summarized as follows:  

\begin{itemize}
\item Despite its continuum-dominated spectrum, \target\ shows clear PAH emission at 3.3, 6.2, and 11.3\um, with a notable PAH power shift favoring the shortest PAH band at 3.3\um. We detect the aliphatic feature at 3.4\um, together with atomic fine-structure lines including \FeII\ (3.21\um) and \OIV\ (25.88\um), as well as various \molH\ transitions (Figure~\ref{fig:tot_spec}). \target\ also exhibits a strong $\Pfdelta$ recombination line at $\sim$3.3\um\ easily detected against the 3.3 PAH emission, in contrast to that seen in typical star-forming galaxies where the PAH feature dominates this part of the spectrum (Figure~\ref{fig:pah33}). The 3.47\um\ broad plateau feature seen in many star-forming systems is mostly absent from the \target\ spectrum, possibly as a result of processing by the hard radiation field.

\item A 3.3\um\ PAH map shows most of the PAH emission in \target\ is concentrated near the northern super star cluster (SSC), spatially coincident with the CO emission detected by ALMA (Figure~\ref{fig:IIZw40_maps}). Molecular gas likely provides shielding to these small PAH molecules in this region. 

\item There is a strong spatial correlation of the \NeIII/\NeII\ and 3.3/11.3 PAH ratios, suggesting a destruction of small grains by UV photons in regions with the hardest radiation fields (Figure~\ref{fig:Ne_pah_ratio}). However, we find \emph{no} evidence that the smallest PAH population is being preferentially destroyed throughout \target. In fact, the 3.3\um\ PAH band appears substantially enhanced compared to longer-wavelength bands, resulting in fractional PAH 3.3 strength (PAH 3.3 / $\Sigma$PAH) approaching 10\% or more (Figure~\ref{fig:pah33_totPAH_Ne}), significantly higher than seen in other star-forming environments. We propose that the PAH size distribution in \target\ is shaped by two mechanisms: photo-processing (top-down) and inhibited growth (bottom-up). In addition, double photon heating might be contributing to the strength of the 3.3 PAH feature in  \target\ due to its extremely high radiation field intensity (log$U\sim$5).

\item The apparent 3.3\um\ PAH flux derived from NIRCam F335M photometry can be strongly affected by the presence of the $\Pfdelta$ emission line and other strong emission lines within the flanking F300M and F360M filters. We find that the photometric method can recover the 3.3\um\ PAH flux within 50\% when the continuum-subtracted F335M surface brightness exceeds 0.1 MJy/sr, but can substantially under-predict the true 3.3\um\ PAH flux at lower surface brightness levels (Figure~\ref{fig:pah33_synphot}). 

\end{itemize}
Future JWST observations of additional BCD galaxies, particularly those with intermediate metallicities similar to \target, will be crucial for establishing a more comprehensive understanding of PAH evolution.

\section{Software and third party data repository citations} \label{sec:cite}

\software{Astropy \citep{Astropy2013, Astropy2018, astropy2022},
          CAFE \citep{Marshall2007, Diaz-Santos2025},
          \emph{JWST} Science Calibration Pipeline \citep{Bushouse2022},
          lmfit \citep{Newville2014},
          Matplotlib \citep{Hunter2007},
          Numpy \citep{VanderWalt2011, Harris2020},
          pyphot \citep{Morgan2025},
          SciPy \citep{Virtanen2020}
          }

\begin{acknowledgments}
The authors thank the anonymous referee for providing useful comments and suggestions. T.S.-Y.L. acknowledges funding support from NASA grant JWST-2511. This work was carried out in part at the Jet Propulsion Laboratory, California Institute of Technology, under a contract with the National Aeronautics and Space Administration. This work was performed in part at the Aspen Center for Physics, which is supported by National Science Foundation grant PHY-2210452. MI acknowledges the support by JSPS KAKENHI Grant Numbers 21K03632 and 25K07359. TN acknowledges the support by JSPS KAKENHI Grant Numbers 23H05441 and 23K17695. 
 
\end{acknowledgments}





%
\facilities{\emph{JWST} (NIRSpec \& MIRI), MAST, NED}

\software{Astropy \citep{Astropy2013, Astropy2018, astropy2022},
          CAFE \citep{Marshall2007},
          \emph{JWST} Science Calibration Pipeline \citep{Bushouse2022},
          lmfit \citep{Newville2014},
          Matplotlib \citep{Hunter2007},
          Numpy \citep{VanderWalt2011, Harris2020},
          SciPy \citep{Virtanen2020}
          }


\appendix
\section{The apparent absence of 7.7, 8.6, and 17\um\ PAH}
\label{sec:no_pah}
\begin{figure*}
    \centering        \includegraphics[width=0.98\textwidth]{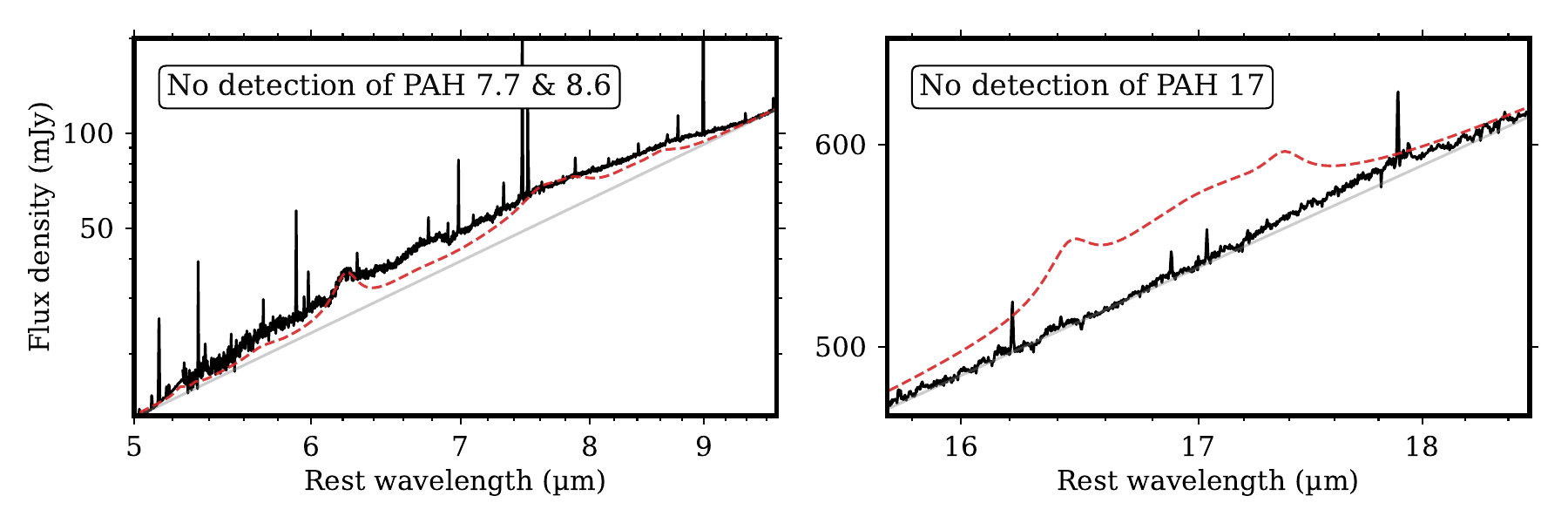} 
    \caption{A zoom-in view of the IFU total extraction shown in Figure~\ref{fig:tot_spec}, focusing on the 7.7 and 17\um\ ranges to demonstrate the non-detection of PAH emission at 7.7, 8.6, and 17\um. Red dashed curves represent the scaled starburst template from \citet{Lai2020}, provided as a reference to illustrate typical PAH profiles in these spectral regions.
    } 
    \label{fig:no_PAH}
\end{figure*}

To illustrate the absence of PAH features at 7.7, 8.6, and 17\um, Figure~\ref{fig:no_PAH} presents spectral zoom-ins of these ranges. The left panel shows the wavelength range between 5.0--9.5\um. For a fair comparison between the observed spectrum and the 1C PAH template from \citet{Lai2020} (as shown in Figure~\ref{fig:tot_spec}), we overlay the template onto a \emph{pseudo-continuum} (gray) defined by a power law based on two pivot points at 5.0 and 9.8\um, wavelengths where PAH emission contributes minimally in this spectral regime. After scaling the template based on the detected 6.2\um\ PAH band, the observed spectrum shows no resemblance to the template at 7.7 and 8.6\um, indicating no detectable emission at these wavelengths. This non-detection may be due to dilution by the strong hot dust continuum. The right panel examines the 17\um\ range based on the same approach, with a power law continuum pivoting at 15.6 and 19.0\um. Despite arbitrary scaling of the PAH profile, the spectrum aligns closely with the continuum (gray), revealing no evidence of the 17\um\ PAH feature.

In Figure~\ref{fig:pah33_totPAH_Ne}, $\Sigma$PAH denotes the sum of the PAH features at 3.3, 6.2, 7.7, 8.6, 11.3, and 17\um, which typically accounts for $>$90\% of the PAH flux in star-forming galaxies. However, in the case of \target, only the 3.3, 6.2, and 11.3\um\ features are robustly detected as discussed in \S\ref{sec:Results}. We define $\Sigma$PAH$_\mathrm{IIZw40}$ to refer specifically to the sum of these three detected features, which we treat as a lower limit for the total PAH emission. To account for the missing contributions from the undetected features, we derive an upper limit on the total PAH flux ($\Sigma$PAH$_\mathrm{UL}$) by estimating the 3$\sigma$ integrated fluxes of the 7.7, 8.6, and 17\um\ bands in the total extracted spectrum. This upper limit is then scaled and applied uniformly across individual spatial cells using a fixed $\Sigma$PAH$_\mathrm{UL}$/$\Sigma$PAH$_\mathrm{IIZw40}$ ratio.

Here, $\sigma$ is defined as the product of the PAH profile's width and the root mean square (RMS) of the residuals after subtracting a 3rd order polynomial fit from the local continuum, where no PAH band and emission line can be seen in the spectrum. Specifically, for the 7.7 and 8.6\um\ features, the RMS is measured over the 9.68---10.43\um\ continuum range, yielding 0.68 mJy. 
For the 17\um\ feature, the RMS derived from the 18.15--18.50 $\mu$m region is 1.93 mJy. The width of each PAH band is defined as the interval spanning one FWHM, centered on the central wavelength reported in \citet{Smith2007}. For PAH complexes comprising multiple sub-features such as the 7.7 and 11.3\um\ bands, the integration range extends from half an FWHM below the bluest sub-feature to half an FWHM above the reddest sub-feature. The resulting ratio of $\Sigma$PAH$_\mathrm{UL}$/$\Sigma$PAH$_\mathrm{IIZw40}$ is 2.81.

\section{Estimate of the radiation field}
\label{sec:radiation_field}
The intensity of interstellar ultraviolet (UV) radiation is often expressed in units of the Habing field (\Gnought). Below, we outline the steps used to derive \Gnought\ for SSC-N.

We adopt the dereddened far-UV (FUV) spectrum of SSC-N from \citet[][Fig.~3]{Leitherer2018}, which shows a relatively smooth continuum between 1200--1700\,\AA. Since this spectral coverage does not extend over the full range required to compute \Gnought, we extrapolate the spectrum linearly to span 912--2400\,\AA\ for the flux integration. The resulting FUV flux is
\[
F_{FUV} = 1.38 \times 10^{-10}~\mathrm{erg~s^{-1}~cm^{-2}}.
\]
Assuming the distance to \target\ of $D$=10\,Mpc, we estimate the total FUV luminosity to be
\[
L_{FUV}=4\pi D^2 F_{FUV} \approx 1.65 \times 10^{42}~\mathrm{erg~s^{-1}}.
\]
Further, assuming the cluster radius of $R$=10\,pc \citep{Leitherer2018}, the local FUV flux at 10\,pc from the cluster can be written as
\[
F_{local}=\frac{L_{FUV}}{4\pi R^2} \approx 138~\mathrm{erg~s^{-1}~cm^{-2}}.
\]
Therefore,
\begin{equation}
    G_{0} = \frac{F_\mathrm{local}}{F_\mathrm{Habing}} = \frac{138}{1.6 \times 10^{-3}} \approx 8.6 \times 10^{4}.
\end{equation}

\noindent Using the relation \( U = 0.88 \times G_0 \) \citep{Draine2007b}, this corresponds to a radiation field parameter of
\[
U = 75{,}680,\quad \text{or} \quad \log U \approx 4.88.
\]

\bibliography{Lai_IIZw40}{}
\bibliographystyle{aasjournalv7}


\end{CJK*}
\end{document}